# Effect of charge transfer band on luminescence properties of Yb doped $Y_2O_3$ nano particles for blue-far yellow emission

**Pratik Deshmukh[a], S. Satapathy[a, *], Anju Ahlawat[a], Khemchand Sahoo[a], and P. K. Gupta[a]**

[a] Nano Functional Materials Laboratory, Laser Materials Development & Devices Division, Raja Ramanna Centre for Advanced Technology, Indore 452013, India

Key words: Photoluminescence; Nanophosphor; Yb:$Y_2O_3$; Charge transfer band; TEM

**Corresponding author**

[*] **E- mail:** srinu73@cat.ernet.in, srinusatapathy@gmail.com  Phone: 91 731 2488660/8658

Fax: 91 731 2488650

**ABSTRACT**

The photoluminescence properties of Yb doped $Y_2O_3$ nanoparticles are investigated in visible region. The presence of peaks at 335 nm and 370 nm in the excitation spectra of bulk as well as in nano Yb: $Y_2O_3$ and their absence in pure $Y_2O_3$ confirm its origin due to $Yb^{3+}$ dopant. The Yb doping in nano $Y_2O_3$ not only modify the charge transfer bands (CTB) but also helps in transition of electron from these CTBs to ground $^2F_{5/2}$ and $^2F_{7/2}$ levels of Yb. This results in strong intense broad emission from ~400 to ~650 nm (white light emission range) for 335 nm excitation.

## 1. Introduction

Charge transfer bands (CTB) are observed in several rare-earth (RE) oxide compounds and these are formed due to the transfer of electronic charge from the highest occupied orbital of the valence band in the host lattice, i. e. from the 2p orbitals of oxygen in the case of oxygen based hosts, to RE ions [1,2]. The CTB transitions are usually observed as strong and broad absorption bands in the UV-visible region as they are not restricted by any selection rules [1]. The CTBs in $Eu^{2+}$ doped $Sr_2SiO_4$ and $Eu^{2+}$ doped $Sr_3SiO_5$ phosphors play an important role in photoluminescence (PL) process for wide range emission in visible region, make these materials suitable for white light emitting diodes[3-5]. The part of CTB absorption energy due to $Eu^{2+}$ in Sr (I) site transfer to $Tb^{3+}$ ion at Sr(II) site and consequently broad luminescence from both $Eu^{2+}$ and $Tb^{3+}$ is observed in $Eu^{2+}$ and $Tb^{3+}$ doped $(Sr,Ba)_2SiO_4$.[6] The broad emission in $Sr_2SiO_4: Eu^{+2}$ is associated with ligand (oxygen) distribution around two types of Sr sites (nine coordination Sr(I) site and ten coordination Sr(II) site). Similarly the rare earth doped $Y_2O_3$ is prone to CTB formation due to oxygen ligand environment around $Y^{3+}$ ions.

In cubic $Y_2O_3$ (space group Ia3) the $Y^{3+}$ ion occupies 24 noncentrosymmetric ($C_2$) and 8 centorsymmetric ($C_{3i}$) sites in an elementary cell. The luminescence of the $Y_2O_3$ is predominantly connected with noncentrosymmetric ($C_2$) site. Since $C_{3i}$ site has an inversion centre, according to the selection rules electric dipole transitions are not allowed for ions occupying this site, because these sites cannot provide odd terms in the crystal field [7]. The $Y_2O_3$ crystallizes in monoclinic phase (space group C2/m) and possesses three crystallographically distinct cation sites with seven fold coordination, each having point group symmetry $C_s$.[8] However the $Y_2O_3$ crystallizes in cubic phase in majority of synthesis method. The $C_2$ and $C_{3i}$ cation sites associated with two different centers of distorted octahedrons which changes the electron distribution between $Y^{3+}$ and oxygen ligands giving rise to two types of CTBs. The $Yb^{3+}$ ion in $Y_2O_3$ occupies any of the two sites i. e. either $C_2$ or $C_{3i}$ sites and hence modifies the electronic distribution in $Y_2O_3$ which in turn changes the energies of CTBs associated with $C_2$ and $C_{3i}$[9]. The Yb doped $Y_2O_3$ (Yb: $Y_2O_3$) is a well studied material for IR laser application (absorption at 940 and 976 nm; lasing action at 1030 and 1070 nm) [10, 11]. However very few reports are available on PL studies of Yb: $Y_2O_3$ in visible region because of low intensity emission due to transition from CTB to ground level. Two broad emission bands are observed at 368 and 533nm in Yb: $Y_2O_3$ for excitation wavelength 215nm at 10K are associated with CT-$^2F_{7/2}$ and CT-$^2F_{5/2}$ transitions but there is no differentiation between the CTBs associated with $C_2$ and $C_{3i}$ centers[12]. The electronic transition arises due to $Yb^{3+}$ in different host materials are extensively studied by L. V. Pieterson[12] and the emission due to $Yb^{2+}$ is extensively reported by Dorenbos[2]. A clear picture on CTBs associated with $C_2$ and $C_{3i}$ centers are still matter of investigation. In the present work the luminescence properties of nano Yb: $Y_2O_3$ was studied at different doping percentage of Yb. The excitation and emission wavelengths of nano Yb: $Y_2O_3$ are correlated using CTB associated with $C_{3i}$ and $C_2$ sites. An energy level diagram of nano Yb: $Y_2O_3$ for visible emission was presented to explain broad emission in visible region. The luminiscence spectra of nano and bulk Yb: $Y_2O_3$ particles were compared and are also compared with bulk $Y_2O_3$. The possible use of nano Yb: $Y_2O_3$ particle due to broad emission in visible range (blue to yellow-green) has also been emphasized.

## 2. Experimental

### 2.1. Synthesis

The Yb:$Y_2O_3$ nano particles were prepared by co-precipitation method[13]. High purity $Y_2O_3$ and $Yb_2O_3$ (Alfa Aesar make, 99.999% purity, product no: 11182) were used as starting materials. The steps involved in preparation of Yb: $Y_2O_3$ nanopowders from its salts are 1) preparation of the Yb doped yttrium nitrate (mother solution), 2) addition of $NH_4OH$ into mother solution till formation of $Y(OH)_3$ precipitate, 3) filtration and washing of precipitate with water, 4) drying and 5) calcinations of precursor powders. The dried Yb doped $Y(OH)_3$ precursor was calcined at 900 °C to get nano powders of Yb: $Y_2O_3$. Using above procedure, Yb: $Y_2O_3$ nano powders were synthesized with 2, 5 and 8 mole percentage of Yb.

*2.2. Characterization*

The phase and crystallite size of powders calcined at 900 °C were determined using X-ray diffractometer (Rigaku X-ray diffractometer with a Ni filtered Cu Kα radiation source and wavelength λ of the source being 1.54 A°). The size and morphology of nanoparticles were determined using transmission microscopy (TEM). Nanoparticles dissolved in chloroform are placed on the grid drop wise and the excess liquid is allowed to evaporate. The grids are examined in Philips CM200 electron microscope equipped with a tungsten cathode and operated at 200kV. The photoluminescence (PL) measurement of nano particles was carried out at different absorption wavelength using PL spectrometer (FLS920-s, Edinburgh Instruments Ltd.) at room temperature. The background excitation source used for this purpose was a 450W ozone-free Xenon arc lamp with continuous output. A calibrated photodiode and a R928P photomultiplier are used as excitation and emission detectors in visible range.

## 3. Results and discussion

The X-ray diffraction patterns of (2-8 mol%)Yb: $Y_2O_3$ nano powders are shown in figure 1. The XRD patterns were simulated and the calculated structural parameters are listed in Table 1. The structure of Yb: $Y_2O_3$ samples was found to be cubic ($Ia\bar{3}$) with slight change in lattice parameter however there is no phase change on Yb doping in $Y_2O_3$[14]. The lattice constant and the lattice volume was decreased with subsequent Yb doping due to smaller ionic radius of $Yb^{3+}$ (0.1004 nm) compared to ionic radius of $Y^{3+}$ (0.104 nm). Debye Scherrer formula was utilized to determine the crystallite size of nano Yb:$Y_2O_3$ and the average size of crystallites are found to be 45-70 nm (table 1). It was found that the crystallite size decreases with increasing Yb content.

The TEM bright field images of nanocrystalline (2-8 mol%) Yb: $Y_2O_3$ are shown in figure 2. The average particle size obtained from TEM micrographs for 2, 5 and 8mole% Yb:$Y_2O_3$ were 65-70 nm, 60-65 nm and 55-45 nm respectively. Particle size was found to decrease with increasing Yb mole percentage in Y2O3 and the morphology of the particles was almost same for all Yb: $Y_2O_3$ nanopowders.

The excitation and emission spectra for 2, 5 and 8% Yb: $Y_2O_3$ nano powders are shown in figure 3, 4 and 5 respectively. The excitation spectra for (2%)Yb: $Y_2O_3$ nano powder was recorded by keeping detector at 420nm and two broad excitation peaks were observed at 335 nm and 370 nm (figure 3(a)). The peak at 370 nm is more intense compared to peak 335 nm for 2%Yb: $Y_2O_3$ however the intensity of the peak at 335 nm is more compared to intensity of the peak at 370 nm (figure 4a) for 5%Yb: $Y_2O_3$. Similarly for 8% doped Yb: $Y_2O_3$ we have observed two excitation bands at 335 and 370 nm like 5%Yb doped $Y_2O_3$ except decrease in intensity of peak at 335 nm (figure 5a). It confirms the existence of two strong bands in between valance and conduction band of Yb: $Y_2O_3$ from which transition of electron takes place at particular excitations i.e. 370 and 335nm. To know the origin of these transition emission in nano Yb: $Y_2O_3$ powder, the PL studies of pure $Y_2O_3$ (bulk powder) and Yb: $Y_2O_3$ (transparent ceramic pellet) were also carried out and are shown in figure 6 and 7 respectively. We observed only one broad peak (maximum at 324nm) in the excitation spectra of bulk $Y_2O_3$ (figure 6(a)) but for Yb: $Y_2O_3$ bulk we observed two excitation peaks at 335 and 370 nm (figure 7(a)) as that of nano Yb: $Y_2O_3$ powders. The presence of 335 nm and 370 nm peaks in the excitation spectra of bulk and as well as of nano Yb: $Y_2O_3$ confirms its origin due to $Yb^{3+}$ dopant.

Figure 3(b) shows the emission spectra of (2%) Yb: $Y_2O_3$ excited at 270, 335 and 370 nm. Broad emission bands at ~359nm, ~400 nm and ~503 nm were observed in emission spectrum of (2%)Yb: $Y_2O_3$ for 270 nm excitation. The broad emission peak at ~503 nm is less intense compared to peak at ~400nm. The broad peak at ~400nm split into two sub peaks centered at 412nm and 432nm. When (2%)Yb: $Y_2O_3$ nano powders are excited at 335 nm, the PL intensity of broad emission peaks i.e. at ~400nm and ~503nm was increased compared to 270nm excitation. The emission spectra of 5%Yb: $Y_2O_3$ excited at different wavelengths is shown in figure 4(b). A broad emission band ranging from 335nm to 600 nm was observed with less PL intensity for 270 nm excitation. The emission spectrum consists of two broad bands centred at ~412 and ~503 nm and extended in a range between ~390nm to 600nm for 335 nm excitation. The energy gap between two bands amounts to ~4500 cm$^{-1}$ which is very less compared to energy difference between $^2F_{5/2}$ and $^2F_{7/2}$ of $Yb^{3+}$. The stoke shifts for ~412 and ~503 bands are 5460 and 9976 cm$^{-1}$ respectively. The broad emission peak at ~412nm appears for all excitation wave length however the broad peak at ~503nm is absent for 370 nm excitation. Figure 5(b) shows emission spectra of 8% Yb: $Y_2O_3$ at excitation wave lengths 270 nm, 335 nm and 370nm. The spectrum is similar to that of 5%Yb: $Y_2O_3$ except decrease in emission peak intensity irrespective of increase in dopant concentration. The decrease in PL intensity for 8%Yb: $Y_2O_3$ compared to 5%Yb: $Y_2O_3$ may be explained on basis of creation of more defect states due to smaller ionic radius of Yb which leads to decrease in lattice parameters. Since more defects states created for 8%Yb: $Y_2O_3$ may leads to high non-radiative relaxation [15] and hence decrease in PL intensity compared to 5%Yb: $Y_2O_3$. It is also reported that the decrease in crystallite size reduces the PL intensity in rare earth doped $Y_2O_3$.[16] In view of the fact that the crystallite size in 8%Yb: $Y_2O_3$ is very small compared to that of 5%Yb: $Y_2O_3$ may be one of the reason of decrease in PL intensity.

In addition to broad peaks very small peaks at 483, 488, 493, 541, 555 and 612 nm are present in PL spectra of Yb: $Y_2O_3$ for 270nm excitation. Small intensity peaks observed at 483, 488, 493 and 612 nm attributed to presence of Yb in $Y_2O_3$.[17] The emergence of 541 and 555 nm peaks in the PL spectrum has been described later in the manuscript.

The emission spectrum of bulk $Y_2O_3$ and bulk Yb: $Y_2O_3$ is shown in figure 6(b) and 7(b) respectively. In bulk $Y_2O_3$ we observed two broad emission peaks at ~410nm and ~468nm for both 335 and 370nm excitations and for bulk Yb: $Y_2O_3$ two broad peaks at ~420nm and ~490nm were also observed for 335 nm excitation. However one broad emission peak at 420nm was observed for 370nm excitation. The presence of high intense broad peak at ~(410-440nm) in all samples with or without Yb doping and in bulk as well as in nano form of Yb: $Y_2O_3$ confirms that this band is associated with stable CTB may be associated with non centrosymmetric $C_2$ sites. The peak position of this band is slightly changed due to Yb doping in $Y_2O_3$ compared to pure $Y_2O_3$.

The broad emission peak observed at ~503nm for nano Yb: $Y_2O_3$ and emission peak at ~490 nm for bulk Yb: $Y_2O_3$ may correspond to a CTB which position depends on the doping percentage of Yb in $Y_2O_3$. So this CTB may be assigned to centrosymmetric $C_{3i}$ centres which appearance depends on size of particle and percentage of dopant in matrix. In $Y_2O_3$ the intensity of peak due to CTB associated $C_{3i}$ centers is very less compared to intensity of the peak associated with $C_2$ centers which may be attributed to very small distortion of centrosymmetricity of $C_{3i}$ centres. In bulk Yb: $Y_2O_3$ the emission peak associated with distorted $C_{3i}$ centres shifted towards higher wave length compared to $Y_2O_3$ and the intensity becomes prominent due to creation of more distortion at $C_{3i}$ sites due to Yb doping. In case of nano Yb: $Y_2O_3$ further distortion occurs at $C_{3i}$ sites to break centrosymmetry and more transitions are allowed so as to observe intense emission at ~503nm. The emission band at ~503nm mainly appears for 335nm or small wave length excitation made it clear that the CTB associated with $C_{3i}$ centers must have higher energy compared to the CTB associated with $C_2$ centers. From above discussion it is confirmed that two types of CTBs associated with $C_2$ and $C_{3i}$ symmetry centers of $Y_2O_3$ are present in Yb: $Y_2O_3$. The position of CTB associated with $C_{3i}$ depends on type of dopant, percentage of dopant and size of the particles etc. These two charge transfer states correspond to ligand interactions at two different $Y^{3+}$ positions in nano $Y_2O_3$ [18]. The separation between these two CTB is around 2823 cm$^{-1}$.

Small intensity peaks are also observed in bulk $Y_2O_3$ (as purchased micron size powder; Alfa Aesar; product no 11182) and bulk Yb: $Y_2O_3$ at 483, 488, 493 and 612 nm for above band gap excitation i. e. 270nm. These are arrtibuted to presence of Yb in the bulk material.[17] The presence of characteristic peaks at 541 and 555 nm confirms the presence of impurity Er in the bulk $Y_2O_3$. Since nano powders of Yb:$Y_2O_3$ are prepared using these $Y_2O_3$ powders small extra characteristic peaks of Er are also present in the PL spectra of Yb:$Y_2O_3$.

The observed excitation and emission spectra has been summarized in the energy level diagram as shown in figure 8. The electronic transition takes place between CTB (associated with $C_2$ and $C_{3i}$ centres of host $Y_2O_3$) and ground level ($^2F_{5/2}$ and $^2F_{7/2}$ created due to Yb doping). Here it worth to note that the 4f-4f transition and CTB-4f are two different types of transitions of the Yb$^{3+}$ ions in crystals. The 4f orbital is shielded from the surroundings by the filled 5s$^2$ and 5p$^6$ orbital. Therefore the influence of host lattice on the optical transitions within the 4f$^n$ configuration is small and 4f-4f transitions place in IR region, while the CTB absorption shows a broad band character and intense than that of 4f-4f transition and lies in visible region.

The excitation peak at 370nm corresponds to transition of electron from ground level ($^2F_{7/2}$) to CTB associated with $C_2$ charge centre and the peak at 335nm corresponds to excitation of electron to CTB associated with $C_{3i}$ charge centre from $^2F_{7/2}$ level because the absorption energy level of $C_{3i}$ centers is at higher energy than that of the $C_2$ centers[19]. Since this transition depends on the number of $C_{3i}$ charge centres at which there is broken of centre of symmetry the intensity of 335nm peak in excitation spectra is different for various percentage doping of Yb in $Y_2O_3$. For 2% Yb doped $Y_2O_3$ nano powder the number of $C_{3i}$ centres for which the centrosymmetricity was broken is less compared to 5% and 8% Yb doped $Y_2O_3$ nano powders.

The possible broad emission peaks ~359 nm, ~400nm (centered at 412 nm &430nm) and ~503nm observed for 270nm excitation correspond to CTB ($C_{3i}$) to $^2F_{7/2}$, CTB ($C_2$) to $^2F_{7/2}$ and CTB ($C_{3i}$) to $^2F_{5/2}$ (~503nm) transitions respectively. According to energy level diagram, for 335nm excitation (i. e. from $^2F_{7/2}$ to CTB ($C_{3i}$)) the possible strong emission bands at ~503nm, 412 nm, 359 nm and 729nm are due to (CTB ($C_{3i}$) to $^2F_{5/2}$), (CTB ($C_2$) to $^2F_{7/2}$), (CTB ($C_{3i}$) to $^2F_{7/2}$) and (CTB ($C_2$) to $^2F_{5/2}$) transition respectively. The peak at 729nm were not observed in PL spectra due to limitation of PL scan range ( to avoid 1st and 2nd order excitation diffraction peak). Similarly for 370nm excitation the strong possible broad emission peak was observed at ~412nm due to electron transition from CTB ($C_2$) to $^2F_{7/2}$. From above discussion it is understood that due to activation of centrosymmetric ($C_{3i}$) sites in nano Yb: $Y_2O_3$ powders strong broad PL ranging from ~400 to ~ 650 nm is observed.

## 4. Conclusions

In the present work we have investigated the photoluminescence properties of nano Yb: $Y_2O_3$ with different doping percentage of Yb in visible region. The nano Yb: $Y_2O_3$ shows intense and broad excitation peak at 335nm and 370nm. The excitation peak at 370nm is obvious due to transition of electrons from $^2F_{7/2}$ to CTB associated with non centrosymmetric $C_2$ centers but the excitation peak at 335 nm is due to transition of electrons from $^2F_{7/2}$ to the CTB associated with distorted centrosymmetric $C_{3i}$ centers. A broad intense emission peak observed at ~503nm is assigned to transition from CTB ($C_{3i}$) to $^2F_{5/2}$ energy levels. The Yb doping in $Y_2O_3$ not only modify the charge transfer bands but also provides ground energy levels to the emitted electron which results in strong intense broad emission ranging from ~400 to ~650 nm. Among various doping percentage of Yb in $Y_2O_3$, the 5%Yb doped $Y_2O_3$ nano phosphor shows highest emission intensity at ~503nm for ~335nm excitation. So by using AlGaN based LED (output peak at ~335 nm with FWHM 10 nm, commercially available) the 5%Yb: $Y_2O_3$ nano phosphor may used to produce white light.

**Acknowledgements** Authors are grateful to Mr. Himanshu Srivastava, ISUD, RRCAT, Indore for TEM of nano particles. We are also thankful to Mr. Gopal Mohod, for his help in infrastructure development for experimental setups.

**Figure Captions:**

1. X-ray diffraction pattern of the Yb: $Y_2O_3$ nano particles; (a) 2%Yb, (b) 5% Yb and (c) 8% Yb. In figure (a) the black line represents experimental, the red circle represents simulated and the difference line is shown below XRD patterns.

2. Transmission electron microscopic images of Yb: $Y_2O_3$ nano powders; (a) 2%Yb, (b) 5% Yb and (c) 8% Yb.

3. Photoluminescent spectra of Yb(2%): $Y_2O_3$;(a) excitation spectra keeping detector at 420nm and (b) emission spectra at excitation 335 nm (Red), 370nm (green), 270nm (black) and 420nm (blue).

4. Photoluminescent spectra of Yb(5%): $Y_2O_3$;(a) excitation spectra keeping detector at 420nm and (b) emission spectra at excitation 335 nm (Red), 370nm (green), 270nm (black).

5. Photoluminescent spectra of Yb(8%): $Y_2O_3$;(a) excitation spectra keeping detector at 420nm and (b) emission spectra at excitation 335 nm (Red), 370nm (green), 270nm (black).

6. Photoluminescent spectra of $Y_2O_3$ bulk powder (a) excitation spectra keeping detector at 420nm and (b) emission spectra at excitation 335 nm (Red), 370nm (green), 270nm (black).

7. Photoluminescent spectra of Yb(1%): $Y_2O_3$ bulk transparent ceramic (a) excitation spectra keeping detector at 420nm and (b) emission spectra at excitation 335 nm (Red), 370nm (green), 270nm (black).

8. Schematic energy level model of Yb: $Y_2O_3$ nano phosphor.

**Table 1:** Structural parameters and crystallite size for nano Yb (2, 5 and 8%): $Y_2O_3$

| Materials name | Lattice Parameters | | | | | | Space group | Volume of Unit Cell($A^3$) | Crystallite size |
|---|---|---|---|---|---|---|---|---|---|
| | a | b | c | α | β | γ | | | |
| Yb: $Y_2O_3$(2%) | 10.595 | 10.595 | 10.595 | 90 | 90 | 90 | Ia-3 | $1.1894*10^3$ | 69.966 |
| Yb: $Y_2O_3$(5%) | 10.592 | 10.592 | 10.592 | 90 | 90 | 90 | Ia-3 | $1.1884*10^3$ | 67.370 |
| Yb: $Y_2O_3$(8%) | 10.588 | 10.588 | 10.588 | 90 | 90 | 90 | Ia-3 | $1.1870*10^3$ | 46.745 |

**Figure 1:**

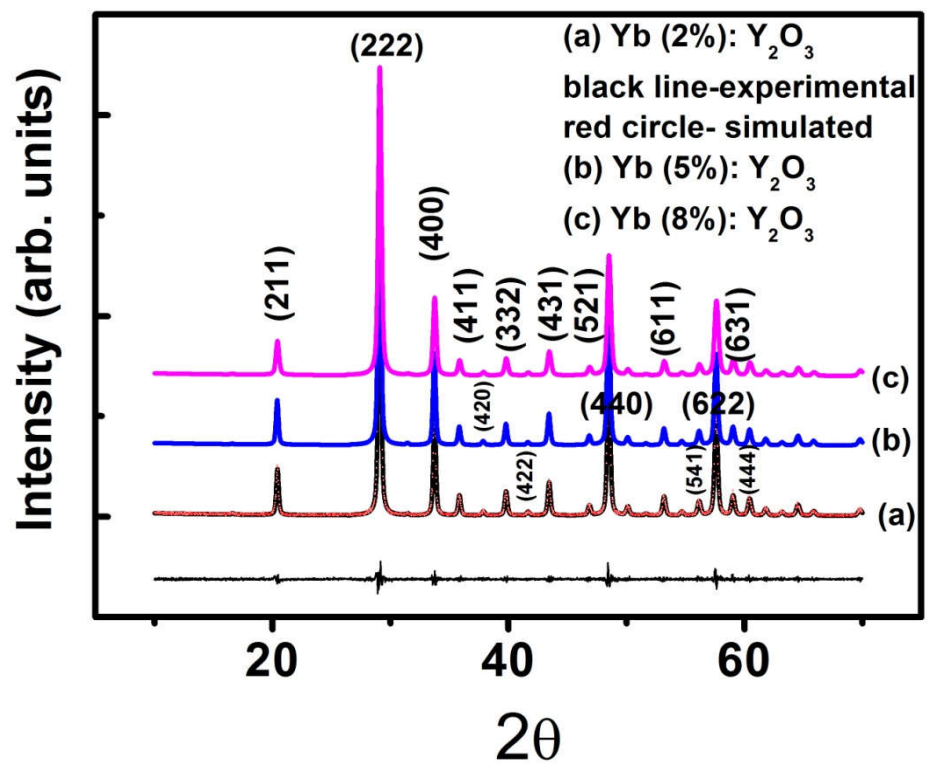

**Figure 2:**

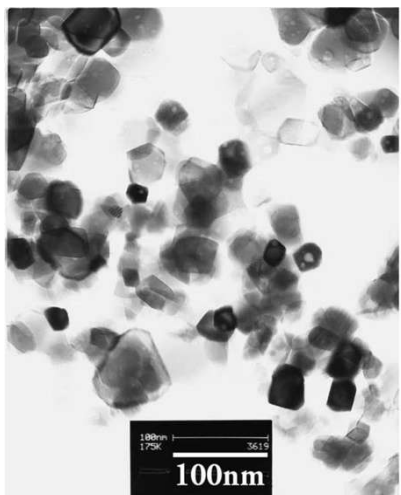 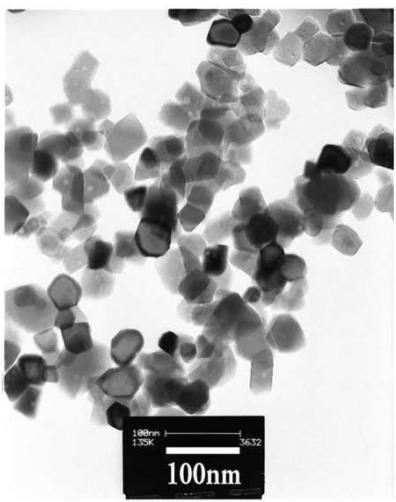 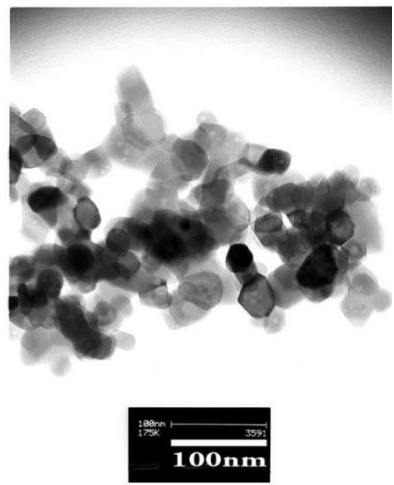

(a) (b) (c)

**Figure 3:**

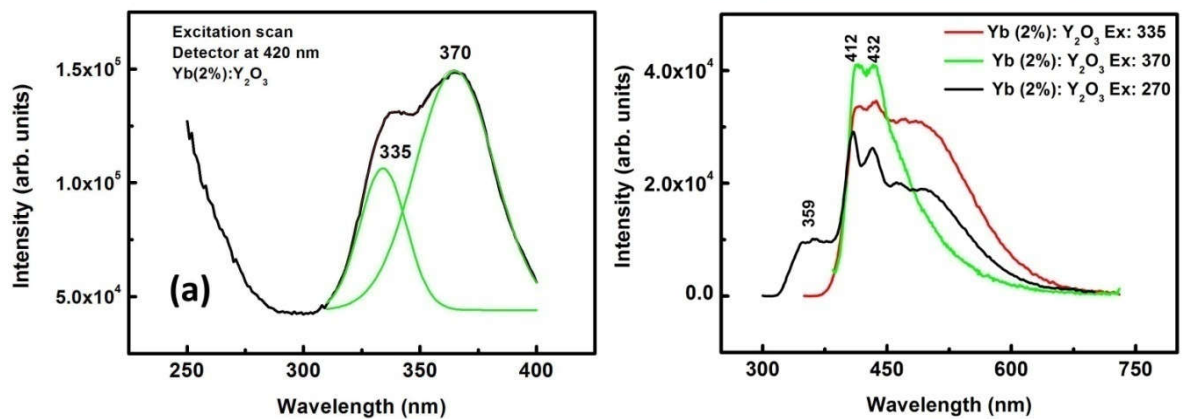

**Fugure 4:**

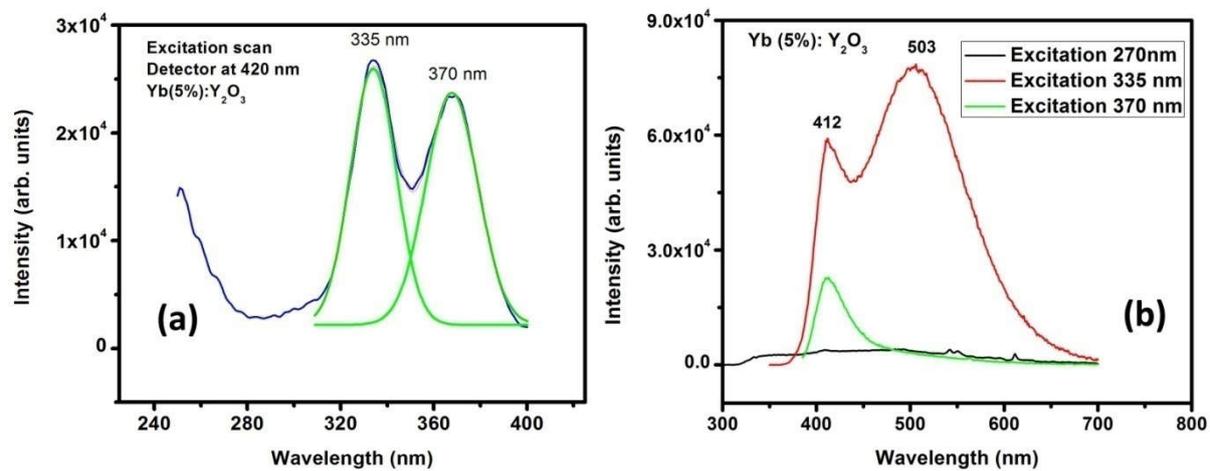

**Figure 5:**

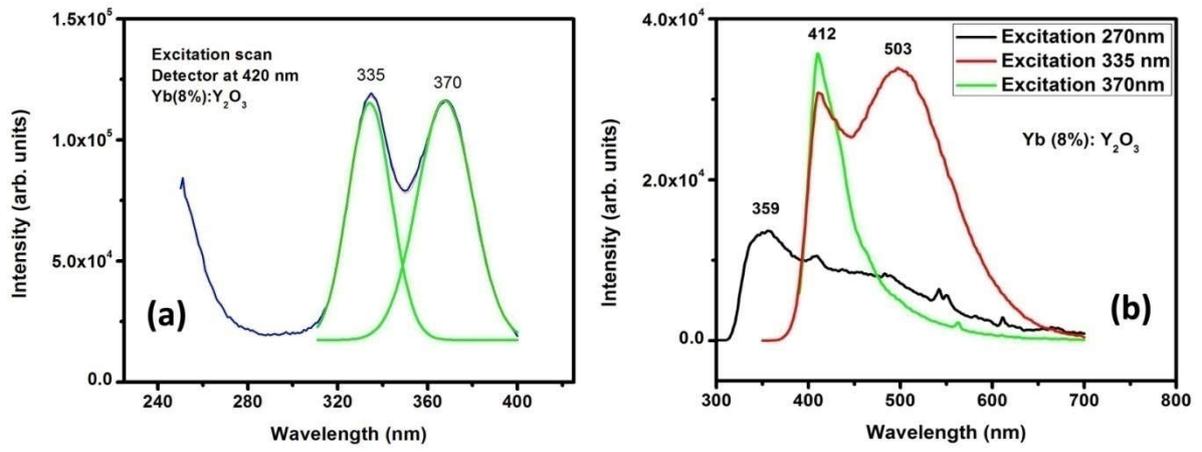

**Figure 6:**

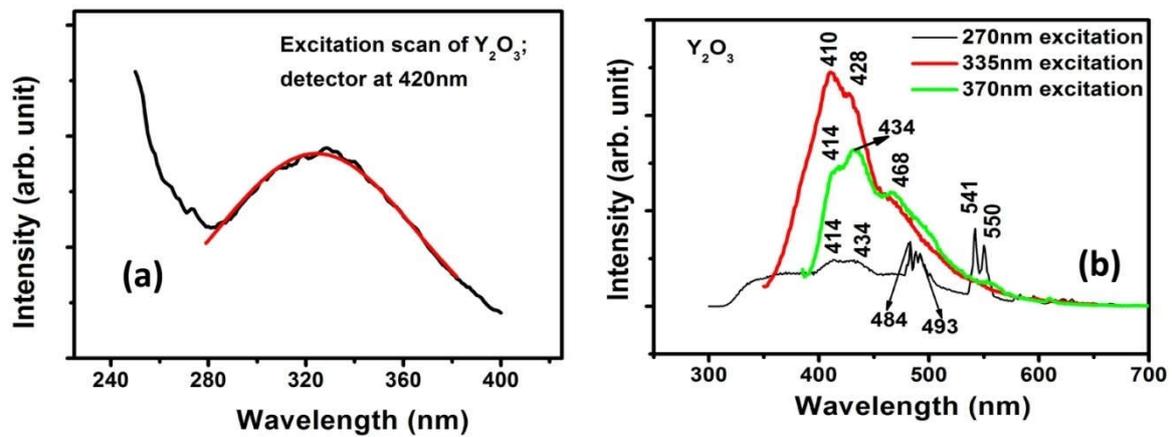

**Figure 7:**

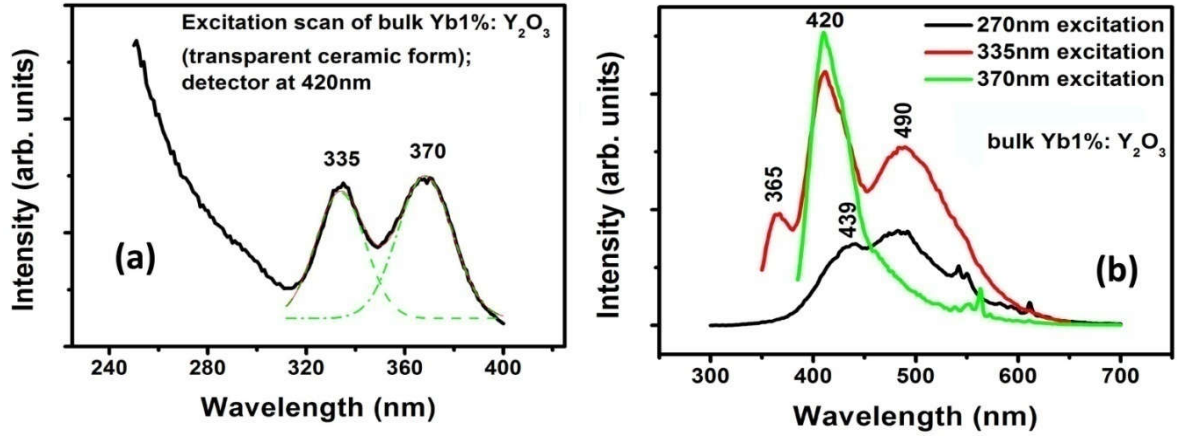

**Figure 8:**

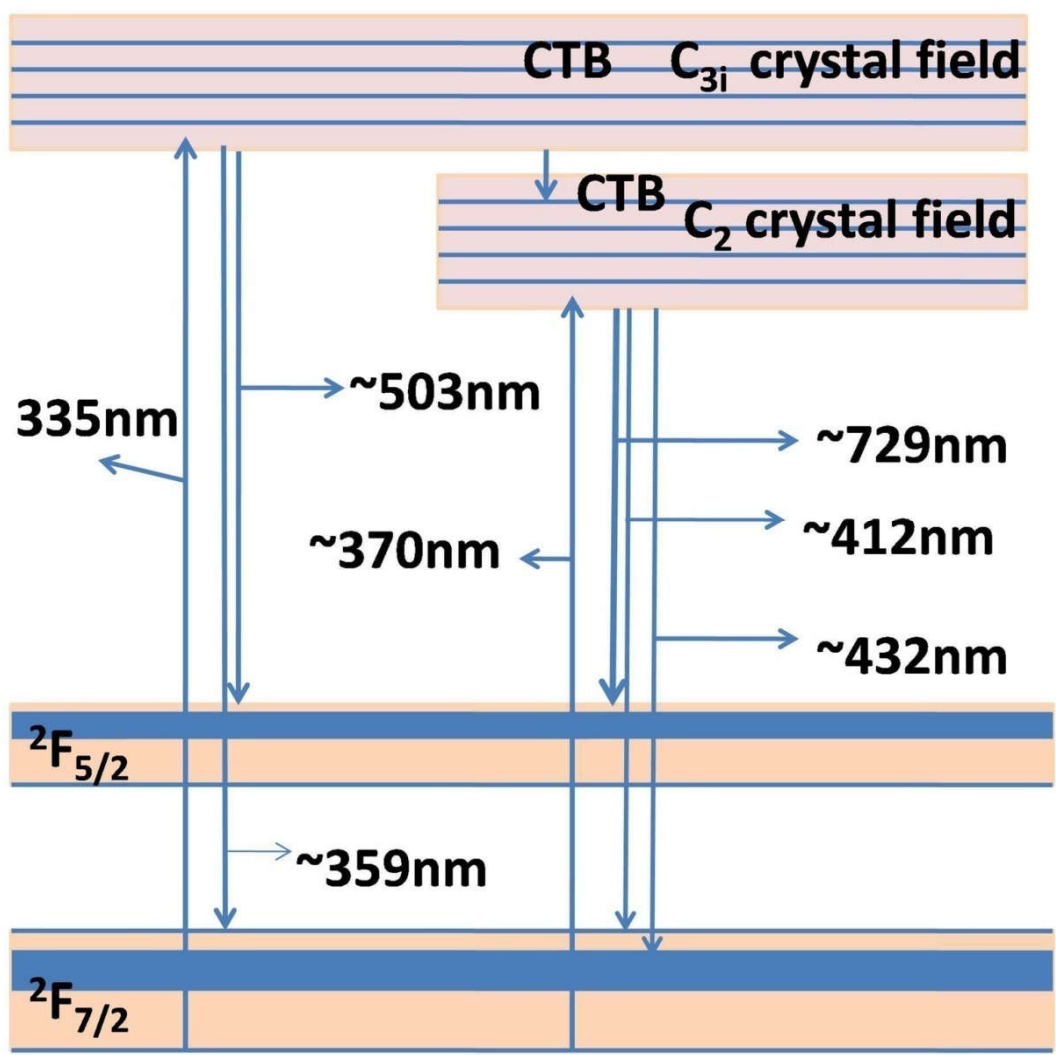